\documentstyle[twocolumn,prb,aps,epsf]{revtex}

\begin{document}
\draft\twocolumn[\hsize\textwidth\columnwidth\hsize\csname 
@twocolumnfalse\endcsname

\title{Quantum phase transitions in the  Bose-Fermi Kondo model}

\author{Gergely Zar{\'a}nd$^{1,2}$ and  Eugene Demler$^1$}

\address{
$^1$Lyman Physics Laboratory, Harvard University, Cambridge, MA \\
$^2$Research Group of the Hungarian Academy of Sciences, Institute 
of Physics, TU Budapest, H-1521 
}
\maketitle

\begin{abstract}
We study quantum phase transitions in the  Bose-Fermi Kondo problem, 
where a local spin is coupled to  independent bosonic and fermionic degrees 
of freedom. Applying a second  order expansion in the anomalous dimension of 
the Bose field we analyze the various non-trivial fixed points of this model. 
We show that anisotropy  in the couplings is relevant at the $SU(2)$ invariant 
non Fermi liquid 
fixed points studied earlier and thus the quantum phase transition is usually 
governed by XY or Ising-type 
fixed points. We furthermore derive an exact result that relates the anomalous 
exponent of the Bose field to that of the susceptibility 
at any finite coupling fixed 
point. Implications on the dynamical mean field approach 
to locally quantum critical 
phase transitions  are also discussed.
\end{abstract}
\maketitle
\date{\today}
\pacs{PACS numbers: 75.20.Hr, 71.10.Hf, 71.27.+a, 72.15.Qm}

]

\narrowtext

\section{Introduction}
Quantum phase transitions in quantum impurity 
models such as the two impurity Kondo 
model,\cite{Jones,Sengupta2imp,Hofstetter,Avishai} anisotropic Kondo models,
\cite{Schlottmann} multi-channel crystal field models\cite{Avi,Koga}
and the Bose-Fermi Kondo model (BFKM) attracted recently a lot of interest. 
These models provide the simplest realizations 
of zero temperature  'quantum phase transitions': They 
have several possible strongly correlated ground states,
each described by some zero temperature ($T=0$) critical point,  
and display a phase transition between these as one 
changes the coupling constants that control their behavior. 
The quantum critical point of these systems is in many cases described by 
an unstable  non-Fermi liquid impurity model with singular thermodynamic 
and transport properties. Recently, it became also possible 
to study some of these quantum phase transitions experimentally in  
mesoscopic devices\cite{Delft}. 

The present paper is devoted to a detailed analysis of the  
Bose-Fermi Kondo problem, consisting of a local impurity spin, 
$S^\alpha$, that couples both to  fermionic ($ \psi$) and bosonic 
($\vec \phi$) fields. In the present paper we restrict our 
discussion to the case
of a spin $S=1/2$ local moment, but our results can be 
trivially generalized for larger spins and many electron 
channels. The bosonic degrees of freedom 
usually represent a fluctuating  magnetic order parameter field 
at a quantum critical point,\cite{Ye,Si96,Si99,SenguptaBFKM,SubirBKM} 
and  exhibit power-law  correlations at zero 
temperature, while $\psi$  represents free Fermions.\cite{footnote}
This model has been proposed in Ref.~\onlinecite{Si} to 
describe the 'local quantum phase  transition' in  alloys like 
$\rm CeCu_{5-x} Au_x$,\cite{Si,Schroder,Piers}. Its 
purely bosonic version  emerges in the context of spin-glass mean field
theories,\cite{Ye,SubirSG} and it has 
been proposed in Ref.~\onlinecite{SubirBKM} 
to describe non-magnetic  impurities  in two-dimensional quantum 
antiferromagnets. 

The  Bose-Fermi Kondo problem in its full version  has been 
analyzed independently by Smith and Si,\cite{Si99} and 
Sengupta\cite{SenguptaBFKM} by performing  a leading  order 
expansion in the anomalous dimension $\epsilon$ of the Bose field, 
$\vec \phi$.  It has been shown in Refs.~\onlinecite{Si99,SenguptaBFKM} 
that the competition  between  the bosonic and  fermionic fields  gives 
rise to  a quantum phase transition, with the  two stable quantum 
phases corresponding  to purely  bosonic and purely fermionic models,
respectively. At the 'fermionic'  fixed point  the impurity spin  
is screened by a Kondo effect, and the 
bosonic field decouples.  At the  bosonic fixed point, on the other hand, 
fermionic fields are irrelevant,  the impurity spin becomes partially 
screened and the properties of this fixed point  are controlled 
by the anomalous  dimension $\epsilon$ of the Bose field.\cite{SubirBKM} 
An expansion to second order in $\epsilon$ 
has been performed for the  isotropic and purely bosonic model 
in Ref.~\onlinecite{SubirBKM}. 

In the present paper, we study the {\em fully anisotropic} Bose-Fermi Kondo 
model, described by the interaction Hamiltonian: 
\begin{eqnarray}
H_{\rm int} &=& H_K + H_B\;, \nonumber\\
H_K^{\rm int}& =& 2\pi \sum_\alpha \lambda_\alpha S^\alpha
(\psi^\dagger {1\over 2} \sigma^\alpha \psi)\;,\nonumber \\
H_B^{\rm int}&= &\sum_\alpha \Lambda^{\epsilon/2} \gamma_\alpha S^\alpha 
\phi^\alpha\;.
\label{eq:H_int}
\end{eqnarray}
Here $\lambda_\alpha$ and $\gamma_\alpha$ ($\alpha=x,y,z$) denote 
dimensionless coupling constants, and $\Lambda$ is a high energy cutoff.
We assume that the fermionic imaginary time propagator corresponds 
to a free  degenerate Fermi gas and  it therefore decays as $\sim 1/\tau$
at $T=0$: 
\begin{equation}
\langle T \psi_\sigma(\tau)  \psi^\dagger_{\sigma'}(0) \rangle =
{1\over 2\pi} {\delta_{\sigma\sigma'}\over \tau},
\label{eq:<psipsi>}
\end{equation}
while the bosonic field  shows critical  imaginary time correlations 
at zero temperature  with an anomalous dimension $\epsilon$
\begin{equation}
\langle T \phi_\alpha(\tau)  \phi_\beta(0) \rangle = cst.\; 
{\delta_{\alpha\beta}\over \tau^{2-\epsilon}}\;.
\label{eq:bose_korr}
\end{equation}
In this work we consider $\epsilon$ as  a given  external 
parameter which will always be taken to be positive. 
In practice, it can be generated by the critical dynamics of  
some spin degrees of freedom at a quantum critical 
point,\cite{SubirBKM,Si} and for 
a dense system of magnetic impurities it must be determined self-
consistently.\cite{Si} 

To analyze the quantum phase transition we carry out  a 
second order expansion  in 
$\epsilon$. As already remarked in Ref.~\onlinecite{SenguptaBFKM},
anisotropy in the couplings $\lambda^\alpha$ 
and $\gamma^\alpha$   is {\em relevant}, even 
in the  purely bosonic model: (a) We find that the
  stable bosonic phase is usually anisotropic, 
and (b) the quantum phase transition in the Bose-Fermi Kondo model 
is in many cases  governed by the new anisotropic fixed points discussed 
below and not the SU(2) invariant fixed point discussed in detail in
Refs.~\onlinecite{SenguptaBFKM,Si99,Si}.
We also  obtain some {\em exact} results for the quantum critical  
behavior that follow from a Ward identity and   are valid  to all orders in 
$\epsilon$. These results parallel the exact results of 
Refs.~\onlinecite{SubirBKM,SubirSG} obtained for the  bosonic versions of the
model, and raise important  questions in the context of the 
theory of Ref.~\onlinecite{Si}. 

Postponing the detailed and rather involved  computations 
to the following sections, let us briefly discuss our main results 
and the methods applied here.  

To describe the quantum phase transitions within the Bose-Fermi Kondo 
model
(BFKM) we used the powerful machinery of multiplicative renormalization group 
(RG). 
The infinitesimal renormalization group  transformations 
are most conveniently described by scaling equations, which  can be obtained 
by performing a  perturbative  expansion 
of the various vertex functions in the couplings $\lambda_\mu$
and $g_\mu \equiv \gamma_\mu^2$, and then gradually eliminating 
high-energy degrees of freedom by reducing the cutoff $\Lambda\to \Lambda'$.
The reduction of the cutoff must be compensated by 
re-scaling the couplings in such a way that the physical 
observables and the singular part of the free energy remain 
invariant. These calculations, detailed in the main body of the paper,
lead to the following Gell-Mann-Low equations:
\begin{eqnarray}
{d\lambda_\alpha\over dl} & =& \beta^{(f)}_\alpha(\{\lambda_\gamma\},
\{g_\gamma\}) \;,\nonumber\\
{dg_\alpha\over dl} &=& 
\beta^{(b)}_\alpha(\{\lambda_\gamma\},
\{g_\gamma\}) \;\label{eq:Gell-mann},
\end{eqnarray}
where $l ={\rm ln}({\Lambda_0\over \Lambda})$ denotes the scaling 
variable with $\Lambda_0$ the initial value of the cut-off, 
 $\beta^{(f)}_\alpha$ and $\beta^{(b)}_\alpha$ 
stand for the fermionic and bosonic beta functions, respectively, 
and we introduced new 'gauge invariant' couplings: 
$$
g_\mu = \gamma_\mu^2\;.
$$
In fact, since the theory is invariant under the $Z_2$  transformation 
$\phi\to -\phi$, physical quantities may only depend on 
$\gamma^2_\mu$.
 
We computed  $\beta^{(f/b)}_\alpha$  to ${\cal O}(\epsilon^3)$ by 
performing a  perturbative  expansion  of the various vertex 
functions in $\lambda_\gamma\sim g_\gamma\sim \epsilon$. 
The explicit expressions are given by Eqs.~(\ref{eq:beta^f_long}) and 
(\ref{eq:beta^b_long}), 
here we only give the leading order results:
\begin{eqnarray}
{d\lambda_x\over dl} & =& \lambda_y \lambda_z - {1\over2} \lambda_x 
(g_y+g_z)\;,
\label{eq:scaling_lambda_leading}
\\
{dg_x\over dl} &=& \epsilon g_x - g_x (g_y + g_z)\;.
\label{eq:scaling_g_leading}
\end{eqnarray}
The other three equations can be obtained by cyclic permutations.

The first term in Eq.~(\ref{eq:scaling_lambda_leading}) corresponds to 
the usual dynamic screening present in the fermionic Kondo model, while 
the linear term in Eq.~(\ref{eq:scaling_g_leading}) reflects the 
anomalous dimension of the bosonic field $\phi$. The last terms in 
Eqs.~(\ref{eq:scaling_lambda_leading}) and (\ref{eq:scaling_g_leading})
are due to the dissipative coupling between the impurity spin and
 the bosons: The bosonic heat bath 
adjusts itself to a given spin configuration and strongly suppresses
processes with spin flips.

Eqs.~(\ref{eq:Gell-mann}) turn out to have a number 
of non-trivial small coupling fixed points with couplings 
$\lambda_\mu,g_\mu \sim \epsilon$: These control 
the possible quantum phases and the transitions between them 
in the BFKM.

\subsection{Non-trivial fixed points}

Before we turn to the general case, it is instructive to 
discuss  the two special cases of (1) purely bosonic couplings
and (2) full $SU(2)$ symmetry with $g_\mu = g$ and
$\lambda_\mu = \lambda$.

\subsubsection{The purely bosonic model}
\begin{figure}[t]
\centerline{\epsfxsize 6cm 
{\epsffile{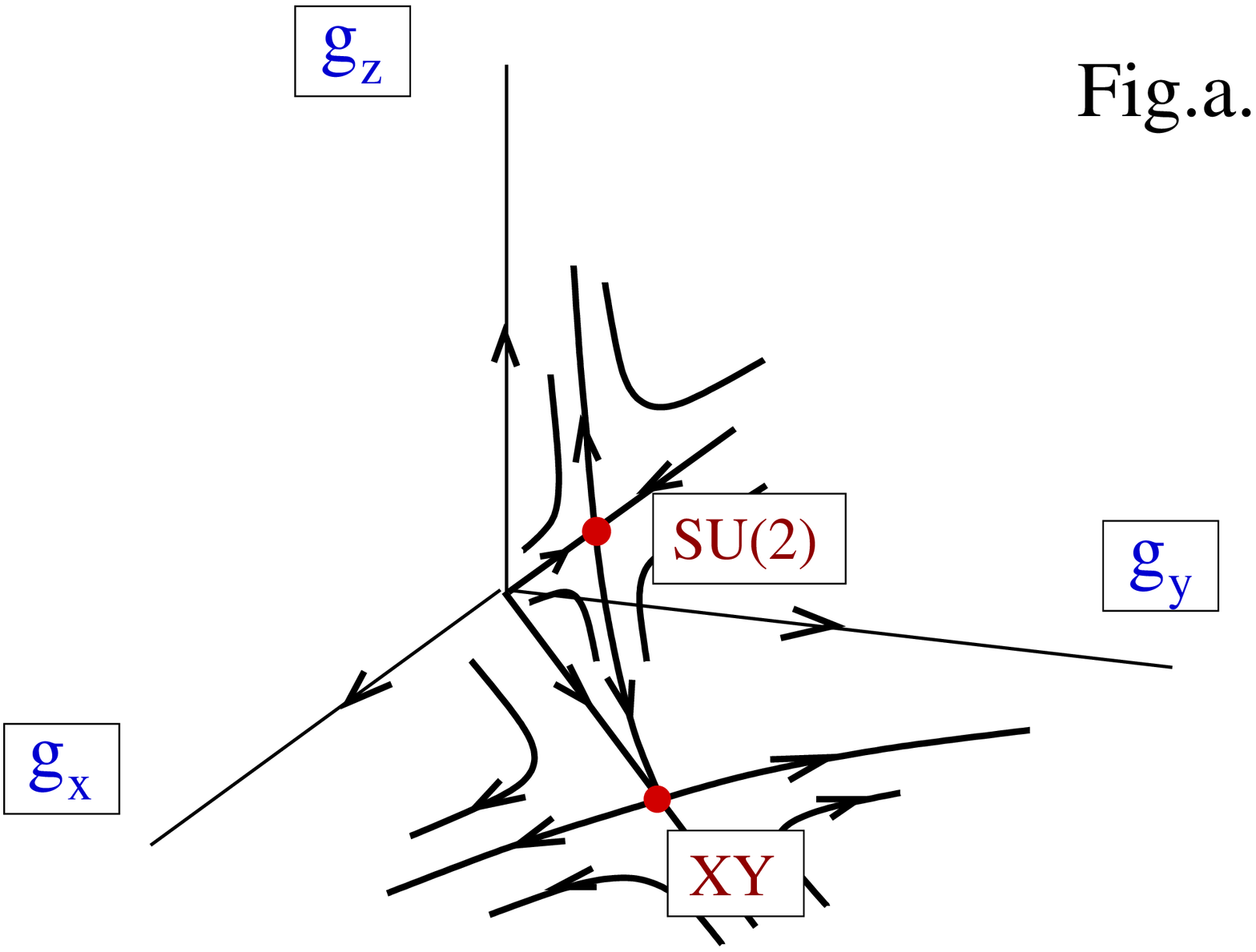}}}
\vspace{0.1cm}
\centerline{\epsfxsize 6cm 
{\epsffile{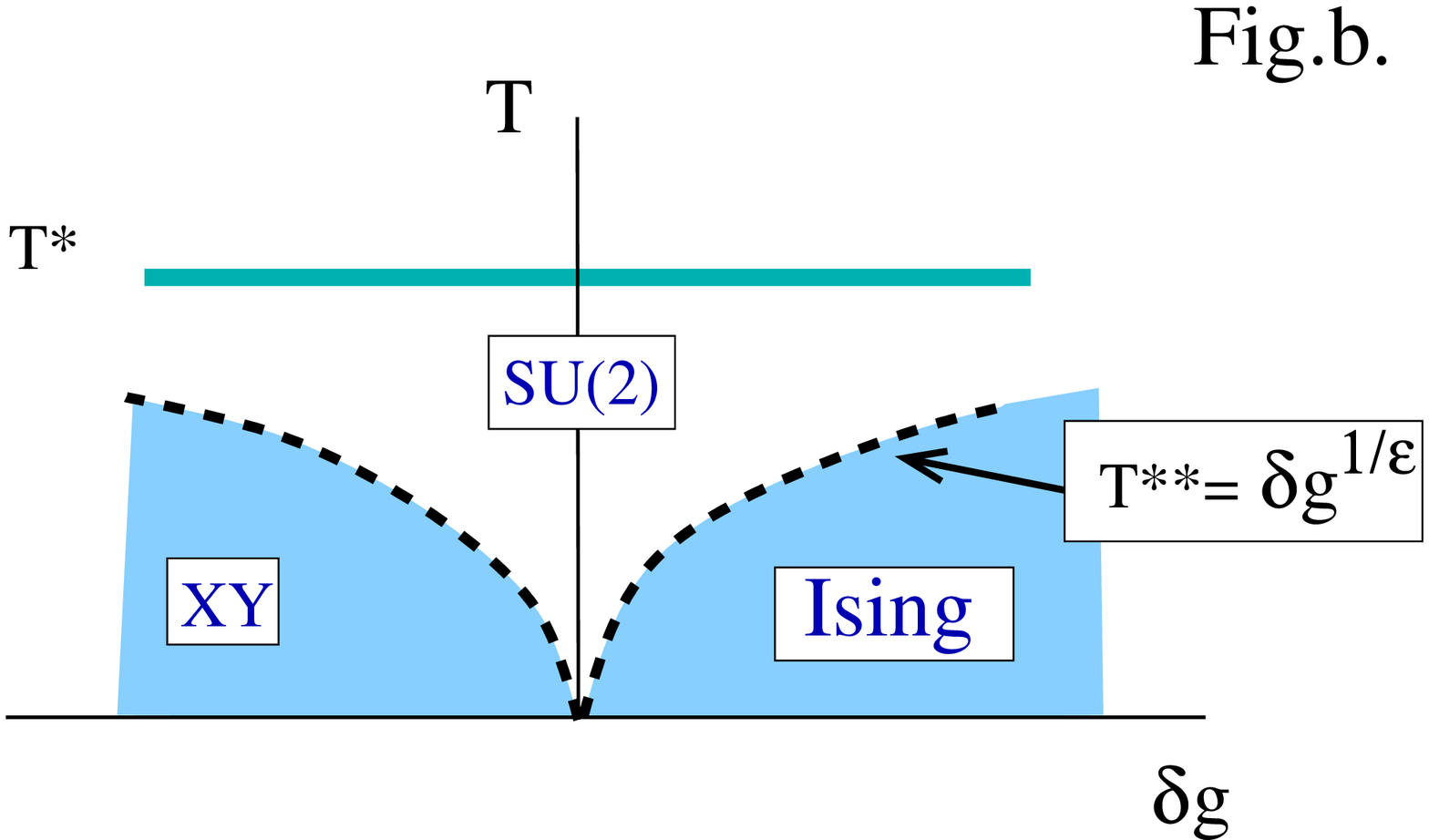}}}
\vspace{0.1cm}
\caption{
\label{bosonic.flow}
Fig. (a) Sketch of the renormalization flows in the purely bosonic model.
We find one non-trivial SU(2)-invariant fixed point and three fixed points
with XY symmetry. Both are unstable 
against breaking the SU(2) and XY symmetries, 
respectively. Fig. (b) Quantum phase transition in the purely bosonic model
with XY symmetry as the anisotropy, $\delta g = g_z-g_\perp$ changes sign. 
The dashed lines indicate cross-over regions corresponding 
to  $T^{**}$.}
\end{figure}

For $\lambda_i = 0$ the RG flows are shown in Fig.~\ref{bosonic.flow}.
Two types of non-trivial fixed points appear:
\vskip5pt
{\parindent = 0pt \em (a) Bosonic SU(2) fixed point} with $g_x = g_y=g_z$:
This is the fixed point analyzed in Ref.~\onlinecite{SubirBKM},
and suggested to control the behavior of non-magnetic  impurities  
in a two-dimensional antiferromagnet  at the 
antiferromagnetic quantum phase transition.\cite{SubirBKM}
\vskip5pt
{\parindent = 0pt \em (b) Three Bosonic XY fixed points} of the type 
$g_x=g_y \ne g_z=0$. 
\vskip5pt 
Both fixed points 
are unstable against spin anisotropy. This instability 
is characteristic to quantum states with finite residual 
entropy:\cite{2CKM,SubirBKM} The  system tries to get rid of this entropy 
by breaking the 
symmetry, and  in fact, the only stable fixed points of the purely bosonic 
model seem to be the three  {\em infinite coupling Ising fixed points}
of the form $g_x=g_y=0\ne g_z\to \infty$.

Even the behavior of the purely bosonic model is already 
very rich. In a realistic case, a number of cross-overs may occur
between  SU(2)-type, XY or Ising behaviors. In particular, 
for a system with XY-type symmetry a quantum phase transition 
occurs  between the XY-type and Ising fixed points 
as $\delta g \equiv g_z- g_x=g_z-g_y$ changes sign 
(see Fig~\ref{bosonic.flow}). 
In this case, the quantum phase transition is controlled by the 
SU(2) type fixed point below an energy scale $T^*$. For small  
$g_\perp = g_x = g_y\approx g_z\equiv g$ one can determine $T^*$  by 
integrating the scaling equations Eq.~(\ref{eq:scaling_g_leading}), and is 
approximately 
given by
$$
T^* \sim \Lambda_0 g^{1/\epsilon}.
$$ 
As $T\to 0$  another cross-over occurs at an energy 
scale 
$$
T^{**}\sim (\delta g)^{y_r}  T^* \;,
$$
where $y_r = \epsilon/2 + \epsilon^2 / 2 + ...$ is the scaling dimension 
of spin-anisotropy at the SU(2) fixed point. Between $T^*$ and $T^{**}$ 
the SU(2)-invariant Bosonic model describes the behavior of the impurity, 
while below $T^{**}$  the physical properties of the model are controlled 
by the  
Ising ($\delta g > 0$) or the XY-type ($\delta g < 0$) 
bosonic fixed points, respectively. 
 
If $g_x \ne g_y$ then the behavior is even more complicated, and eventually 
two cross-overs may occur in series, corresponding to the 
$SU(2)\to XY$ and $XY\to Ising$ transitions. The cross-over regions 
are governed in this case by the SU(2) and  XY-type fixed points, 
respectively.

Although anisotropy is relevant at both ${\cal O}(\epsilon)$ 
fixed points above, 
in many cases the SU(2)-invariant or XY-type fixed points can be also of 
physical interest: In the case of couprates, {\em e.g.},\cite{SubirBKM}
 spin-orbit interaction
is weak and therefore the anisotropy is presumably weak. In this 
case the SU(2)-symmetrical fixed point may appropriately  describe 
the physics over a wide energy range of interest. On the other hand, 
in most magnetic heavy Fermion compounds spin-spin interactions 
are generically  anisotropic,\cite{2CKM} 
and the physics is controlled by XY type or Ising fixed points.

\subsubsection{The SU(2) symmetric Bose-Fermi Kondo Model}

\begin{figure}[t]
\centerline{\epsfysize 5cm 
{\epsffile{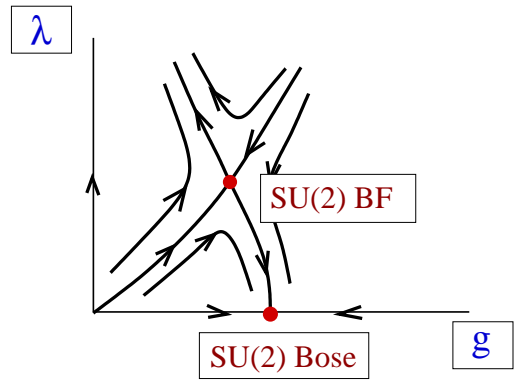}}}
\vspace{0.1cm}
\caption{
\label{fig:bk_flows} Renormalization group flows for the 
SU(2) symmetrical Bose-Fermi Kondo model. ``BF'' denotes the SU(2) symmetrical 
Bose-Fermi fixed point. Here the spin is partially screened and both the 
Fermi and Bose fields couple to it. At the Bose fixed point the fermionic 
degrees of freedom fully decouple from the spin.
}
\end{figure}

For the sake of clarity, we first discuss   the renormalization 
group flows in the $SU(2)$ symmetric case 
with $g_x=g_y=g_z\equiv g$ and $\lambda_x=\lambda_y=\lambda_z\equiv \lambda$.
The results we obtain to second order in $\epsilon$ are qualitatively the same 
as the leading order ones derived in Refs.~\onlinecite{Si99,SenguptaBFKM}. The flows
are sketched in Fig.~\ref{fig:bk_flows}. Two stable quantum phases appear:
\vskip5pt
{\parindent 0pt \em (a) The Kondo state ($\lambda\to \infty$, $g=0$):} This is 
the familiar Kondo state.\cite{Hewson} 
The local moment is completely screened by the Fermions and forms 
a Fermi liquid. It is thus fully  decoupled from the bosonic field. 
At the Kondo fixed point both anisotropy and the bosonic field are irrelevant.
\vskip5pt
{\parindent 0pt \em (b) Purely bosonic SU(2) invariant fixed 
point ($g\approx \epsilon/2$, $\lambda=0$):} 
This is the same fixed point as the one 
we discussed in the previous subsection. 
At this fixed point the coupling to the Fermions is irrelevant, however, 
breaking the spin anisotropy, as discussed earlier, is a relevant perturbation.
It is though a stable fixed point if for some reason exact SU(2) 
symmetry  is guaranteed, 
however, it is generally unstable in non-cubic systems with spin-orbit 
interaction. At this fixed point the impurity is only partially 
screened.\cite{SubirBKM}

These two phases are separated by an interesting quantum critical point: 
\vskip5pt
{\parindent 0pt \em (c) The SU(2) symmetrical Bose-Fermi} fixed point 
($g\sim \lambda \sim \epsilon$). This fixed point governs the quantum 
phase transition between the bosonic and Fermionic SU(2) invariant phases in 
case of 
full SU(2) symmetry. However, similar to the SU(2) invariant Bose fixed point, 
anisotropy is {\em relevant} at this fixed point (in fact, it is the leading 
relevant 
operator, {\em i.e.}, it is more relevant than the operator corresponding to 
the quantum phase transition.) Therefore, in general, it is not this 
fixed point  that controls the quantum phase  transition.

\subsubsection{The general case: Anisotropic BFKM}
The general anisotropic Bose-Fermi Kondo model has  all the fixed points 
discussed above. In addition, however, we find two new types of fixed 
points with XY  and Ising symmetry:
\vskip5pt
{\parindent 0pt \em (a) The XY Bose-Fermi fixed point of the type 
$\lambda_x=\lambda_y\ne \lambda_z$ and $g_x = g_y \ne g_z =0$:}
Similar to the SU(2) Bose-Fermi fixed point, this fixed point separates the  
XY or Ising bosonic phases and the SU(2) invariant Kondo phase, 
and controls
the quantum phase transition between them in case of XY symmetry. \\

Breaking 
the SU(2) symmetry is  a relevant perturbation at the XY-Bose-Fermi 
 fixed points 
too: It corresponds to an operator within the critical surface  
separating the bosonic and fermionic quantum phases. 
If we break the XY symmetry in this direction, the RG flows 
stay within the critical surface and end up in a new 
strong coupling fixed point: 
\vskip5pt
{\parindent 0pt \em (b)} {\em The Ising Bose-Fermi}
 fixed point with structure, $\lambda_x=\lambda_y\ne \lambda_z$ and 
$g_x = g_y=0 \ne g_z$, with $g_z, \lambda_z\sim 1$. 
\vskip5pt
This fixed point is outside the region of applicability of
our perturbative RG. We believe, however,  that this strong coupling 
fixed point 
is not a mere artifact of our approximations, and that this is the 
fixed point
that should describe the phase transition in case of Ising symmetry.

\begin{figure}[t]
\centerline{\epsfxsize 7.1cm 
{\epsffile{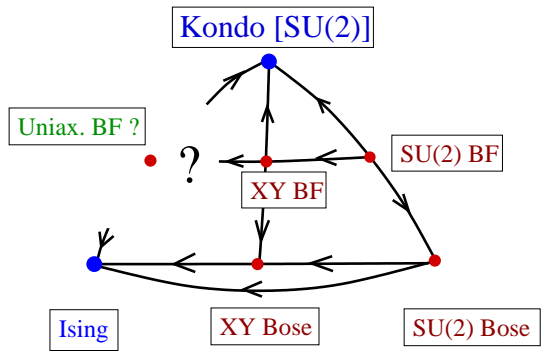}}}
\vspace{0.1cm}
\caption{Summary of various fixed points and the possible flow 
directions. 
}
\label{fig:summary}
\end{figure}

The complicated structure of the flows and the connection between 
the various fixed points  is summarized in Fig.~\ref{fig:summary}.

\subsection{Physical quantities}
Within  the $\epsilon$-expansion, we can compute various physical 
quantities. The scaling properties of these are determined by the scaling 
dimensions and structure of the various relevant and irrelevant 
operators. Instead of enumerating all of these at each fixed point 
we rather list only the most important scaling dimensions  in Table~\ref{fptable}. 

At the ${\cal O}(\epsilon)$ fixed points we enumerated three scaling 
dimensions: \\
(a) The {\em leading relevant operator}: This is usually associated with 
{\em breaking the symmetry} of the fixed point. \\
(b) We also gave the scaling 
dimension of the {\em most relevant operator respecting the symmetry} of the 
fixed point. This operator at the Bose-Fermi fixed points
corresponds to the {\em quantum phase transition} 
between the bosonic and fermionic fixed points with that symmetry. \\ 
(c) Finally, we also enumerated the scaling dimension of {\em 
leading irrelevant} operators,
which  determine the scaling properties of physical quantities 
{\em at} the quantum critical point (or the symmetry-stabilized XY or 
SU(2) Bose fixed points).

\subsubsection{Susceptibility} 
Magnetic field is a relevant perturbation in all non-trivial 
${\cal O}(\epsilon)$ fixed points. To study its effect we added
the following term to the Hamiltonian:
\begin{equation}
H_m = - \sum_{\alpha=x,..,z} h_\alpha S_\alpha  \;,
\label{eq:h_m}
\end{equation}
where $h_\alpha$ denote the components of a local magnetic field. 
As we shall prove later, the scaling equations of the dimensionless 
magnetic field, ${\tilde h}_\alpha \equiv h_\alpha/\Lambda$, are determined by 
the same 
beta-functions as those of the vertex $g_\alpha$:
\begin{equation}
{d {\rm ln}{\tilde  h}_\alpha\over dl} = 1-{\epsilon\over 2}
 + {1\over 2 g_\alpha} \beta^{(b)}_\alpha (\{\lambda_\gamma\},
\{g_\gamma\}) + {\cal O}({\tilde  h})\;.
\label{eq:h_scaling}
\end{equation}
This remarkable result, which is
very similar to the one obtained in Ref.~\onlinecite{SubirSG}
and follows from a Ward identity, has  important 
consequences: By Eq.~(\ref{eq:Gell-mann}) it 
implies that at a fixed point
with coupling $g_\alpha^*\ne 0$,  the 
scaling dimension of the $\alpha$ component of the 
local magnetic field only depends on the 
anomalous dimension of the bosonic bath, and is  {\em exactly}
given by $y_{h,\alpha} = 1-\epsilon/2$. 

As explained in Ref.~\onlinecite{Cardy},{\em e.g.}, 
the dimension $x_{\cal O}$ 
of a  scaling operator ${\cal O}$ at a non-trivial fixed point 
is defined through the asymptotic behavior of its correlation function:
$\langle {\cal O}({\bf r}) {\cal O}({\bf 0})\rangle
\sim 1/r^{2\; x_{\cal O}}$, where ${\bf r}$ denotes the coordinate
variable. The dimension $x_{\cal O}$ is directly related to the scaling
dimension $y_{H}$ of the dimensionless field $H$ that couples to the 
the scaling operator: Since in a $D$-dimensional theory  the 
free energy has a dimension $D$,  $y_{H} = D-x_{\cal O}$. 
In our case the impurity spin lives in an effective 
$D=0+1$-dimensional space, therefore this relation 
immediately relates the scaling dimension of the magnetic 
field $h$ to that of the impurity spin $S$ as $x_S = 1- y_h$.
As a consequence,  the  spin correlation function at 
the SU(2)-invariant Bose and Bose-Fermi fixed points decays at
$T=0$  temperature  as 
\begin{eqnarray} 
\chi^{SU(2)}_{\alpha\beta} (\tau)& =& \langle T S_\alpha(\tau) 
S_\beta(0)\rangle = \delta_{\alpha\beta}\chi(\tau)\;,
\\
\chi(\tau)  & =&  {1\over (T^* \tau)^\epsilon}
\label{chi_tau}
\end{eqnarray}
with $T^*$, a dynamically generated energy scale similar to the 
Kondo scale. This result has been derived earlier for 
the SU(2)-invariant bosonic fixed point in Ref.~[\onlinecite{SubirBKM}].

At the XY fixed points, however, the scaling properties of the magnetic 
field depend on its direction. To be specific, let us assume
that the value of the couplings at the fixed point is 
$g_x^* = g_y^* \ne g_z^* = 0$. In this case,  
spin correlations of the $x$ and $y$ spin components
 decay at $T=0$ as 
\begin{equation}
\chi^{(XY)}_\perp (\tau) =  {1\over (T_\perp^* \tau)^\epsilon}\;,
\label{chi_perp_tau}
\end{equation}
while the decay of correlations in the $z$-direction
is fixed point specific:
\begin{equation}
\chi^{(XY)}_{z}(\tau) =  {1\over (T_{z}^* \tau)^{\eta_{z}}}\;,
\label{chi_par_tau}
\end{equation}
where, up to second order in $\epsilon$,  $\eta_{z}$ is given by 
$\eta_z = 2\epsilon + \epsilon^2$ and $\eta_z= 
2\epsilon + 7\epsilon^2/4$ 
for the XY Bose and XY Bose-Fermi Kondo fixed points, respectively. 
Note that Eqs.~(\ref{chi_tau}) and (\ref{chi_perp_tau}) 
are valid to {\em all orders in $\epsilon$}, while the 
exponents $\eta_{z}$ can be computed only approximately.

The results above apply only at $T=0$ temperature. On  can, however, 
obtain the finite temperature form of the correlation functions and 
their Fourier transforms assuming conformal invariance at these
non-trivial fixed points.\cite{Tsvelik} For the imaginary part 
of the Fourier transform of the local spin-spin correlation functions, 
measurable through neutron scattering experiments, we obtain the 
following scaling form:
\begin{equation}
\chi^{''}_{\alpha}= \Bigl[{ T^*  \over T}\Bigr]^{1-\eta_\alpha}
f(\eta_\alpha,{\omega\over T}) \; {\rm sgn} \omega 
\;,
\label{eq:scaling_form}
\end{equation}
where the scaling function $f(\eta,x)$ behaves as
\begin{equation}
f(\eta,x) \approx 
\left\{
\begin{tabular}{lr}
$  C_>\; |x|^{\eta-1}$ &  \phantom{nn} $|\omega|\gg T$, \\
$ C_<\; |x|$ & \phantom{nn} $|\omega|\ll T$.
\end{tabular}
\right.
\end{equation}
The function $f(\eta,x)$ can be expressed analytically from 
the exact  expression  Eq.~(\ref{chi_full}).

\subsubsection{Resistivity}

Here we only discuss the impurity resistivity 
{\em at the quantum critical points}, {\em i.e.}, we assume 
that all relevant 
variables and symmetry breaking terms 
have been tuned to zero or vanish for reasons of symmetry. 
In this case, the temperature
dependence of the resistivity is
determined by the scaling dimension $y_i<0$ of the leading 
irrelevant operator.
We enumerated this latter for the various fixed points in 
Table~\ref{fptable}. 

The simplest way to obtain the resistivity is to compute the 
impurity scattering rate. This turns out to be proportional to 
$ \sum_\alpha \lambda^2_\alpha(T)$, where 
$\lambda_\alpha(T)$ is the effective coupling at energy scale 
$\omega \sim T$  that can be computed 
from the scaling equations. Since at the bosonic
fixed point the effective couplings $\lambda_\alpha$ vanish
as $\sim \lambda_\alpha(T) \sim T^\Delta$ with a power 
$\Delta = - y_i$ the resistivity there also vanishes as 
\begin{equation}
R_{\rm imp}^{\rm Bose}(T) \sim A \; T^{2\Delta}\;,
\end{equation}
and the impurity fully decouples from the electrons at $T=0$.

The situation at the SU(2) and XY Bose-Fermi fixed points is, 
on the other hand,  different. There the couplings scale to 
a {\em finite} value, corresponding 
to a finite scattering rate, and the leading corrections to this 
scale as
\begin{equation}
R_{\rm imp}^{\rm BF}(T) \sim A + B\;  T^{\Delta}\;,
\end{equation}
similar to the multichannel Kondo model.

The above results only apply for the case of a dilute system
of independent impurities. For the dense systems the interaction
between individual impurities must be taken into account
which leads in general to a different temperature dependence. 

\subsection{Implications for a dynamical mean field theory 
of locally quantum critical phase transitions}

It has been shown in a beautiful series of neutron scattering measurements by 
Schr\"oder {\em et al.} that the quantum critical behavior in $\rm CeCu_{5-x} 
Au_x$ is entirely local within experimental accuracy. This material shows a 
quantum phase transition from a non-magnetic heavy Fermi liquid state to a 
metallic antiferromagnetic state under $\rm Au$ doping. 
The  Fermi liquid energy scale below which 
the Ce spins are screened appears to go to zero as one approaches the 
quantum critical point from the metallic side, indicating that 
at the quantum critical point (QCP) spins remain asymptotically free
at low temperature.\cite{Piers} Schr\"oder {\em et al.}
also find that  at the quantum critical point the dynamic 
susceptibility behaves with a good accuracy as
\begin{equation}
\chi^{-1}(\omega,T,{\bf q}) \sim
{\rm cst.} (\chi^{-1}({\bf q}) + (T - i\omega)^\alpha)
\label{eq:c_exp}
\end{equation}
with an exponent $\alpha \approx 0.75$, where $\chi^{-1}({\bf q})$ 
is an approximately frequency and temperature independent 
function with minima at the ordering wave vectors.

Clearly, the quantum phase transition in $\rm CeCu_{5-x} Au_x$ is driven 
by the 
competition between two mechanisms with which the system can 
get rid of the residual entropy of the Ce impurity spins: 
The Kondo effect that screens magnetic impurities 
individually, and the magnetic ordering, which tends to align 
spins and thus destroys the Kondo effect.

Expression (\ref{eq:c_exp})  is consistent with the presence of 
mean-field-like spatial correlations but simultaneous
 non-trivial local correlations in the time direction.
Based on these observations,\cite{Piers} Si {\em et al.} proposed 
a self-consistent version of the Bose-Fermi Kondo model\cite{Si} 
 as a good candidate to  describe the above quantum critical 
behavior within a dynamical mean field approach.\cite{Kotliar}

\subsubsection{SU(2) invariant theory}

In their work, 
Si {\em et al.} assumed that the system is $SU(2)$ symmetrical
in spin space, and therefore 
identified the QCP with the SU(2) invariant Bose-Fermi 
fixed point. In this theory, the exponent $\epsilon$ must be determined 
self-consistently, and it is  fixed by the 
condition that the correlation functions 
$\langle T S_\alpha (\tau) S_\alpha (0)  \rangle$ and 
$\langle T \phi_\alpha (\tau) \phi_\alpha (0)  \rangle$
must decay asymptotically with the same exponent. 

As we have discussed above, at the SU(2) invariant Bose-Fermi 
fixed point $\langle {\bf S} (\tau) {\bf S} (0)  \rangle \sim 
{1/\tau^{\epsilon}}$ as a consequence of a Ward identity.
Similar to the case of spin glasses,\cite{Ye,SubirSG} 
this immediately determines  the only possible exponent:
\begin{equation}
\epsilon_{\rm QCP}^{\rm SU(2)} \equiv 1\;.
\end{equation}
This result has dramatic consequences: 
\vskip5pt
{\parindent=0pt (a)} It implies that - if a stable self-consistent 
solution of the extended dynamical mean field theory exists - 
then local  
spin correlations decay as $1/\tau$ at the quantum critical point and thus 
the the local  susceptibility must diverge 
logarithmically as a function of frequency or temperature. 
This may be consistent  with the mean field expression 
(\ref{eq:c_exp})   only if the magnetic fluctuations are 
approximately {\em two-dimensional}.\cite{Si}  In this case the exponent
$\alpha$ would be non-universal and would depend on the specific 
properties of the model. 
\vskip5pt
{\parindent=0pt (b)} As a consequence, the scaling dimension of the local magnetic field
(i.e. a field applied only to the impurity) would be {\em exactly} 
$y_h^{\rm SU(2)} =1/2$.
This in turn immediately implies an $h^2/T$ scaling, which is clearly 
in disagreement with the experimental data, showing an almost perfect 
$h/T$ scaling. However, in the experiments one applies 
a {\em global} field, and we know  cases where the  
the scaling dimension of the global field is different from that of the 
local field acting only on the impurity.\cite{SubirBKM,Kevin} 
It is therefore conceivable that the global field could result in 
an $h/T$ scaling.

All these considerations are only valid at the SU(2) invariant 
Bose-Fermi Kondo fixed point. However, magnetic interactions
in $\rm CeCu_{5-x} Au_x$ and most heavy fermion materials
are strongly spin-anisotropic, and we have shown that anisotropy 
is relevant at the $SU(2)$-invariant Bose-Fermi Kondo fixed point.

Thus if there is any locally critical quantum field theory that 
describes the QCP, it {\em must} be anisotropic. 
One of the candidates for the non-trivial fixed points is the 
new XY Bose-Fermi fixed point. At this fixed point, 
the coupling to the  $z$-component of the spin is irrelevant 
and scales to 0. This is therefore, a good candidate to describe 
systems like $\rm YbRh_2 Si_{2-x} Ge_x$,\cite{Piers}, 
where magnetic fluctuations are more of XY-type.

In $\rm CeCu_{5-x} Au_x$, however, the magnetic degrees of freedom are 
more Ising-like. The corresponding quantum phase transition 
can therefore possibly be described by the Ising Bose-Fermi fixed
point of the local model with structure $g_x = g_y=0$, $g_z\sim 1 $, and 
$\lambda_z \ne \lambda_x=\lambda_y$.

\subsubsection{XY invariant theory}

As we discussed earlier, at this fixed point  correlations in the $xy$ and $z$ 
spin directions decay with different powers. This immediately implies that 
in a self-consistent theory one must also assume that the $xy$ and $z$ 
components of the field $\phi_i$ have also different  anomalous dimensions, 
$\epsilon_{\perp}$ and $\epsilon_z$. It is straightforward to generalize our 
computations to this case. Similar to the $\epsilon_{\perp} = \epsilon_{z}$ 
case, the asymptotic behavior of the correlations of the $xy$ 
spin components is 
determined by a Ward identity, and is given by $\langle S_x(\tau)S_x(0)\rangle 
\sim \langle S_y(\tau)S_y(0)\rangle \sim 1/\tau^{\epsilon_{\perp}}$, which 
yields again  the exact relations,
\begin{eqnarray}
\epsilon_{\perp, \rm QCP}^{XY} \equiv 1\;,\\
y_{h\; \perp}^{\rm XY} =1/2\;.
\end{eqnarray}
and implies conclusions similar to the ones drawn for 
the SU(2)-invariant fixed  point for spin correlations within the 
$xy$ plane.

The other exponent, $\eta_z$, is however, not determined by a Ward identity
And therefore the condition $\eta_z (\epsilon_z)= 2-\epsilon_z$ yields a non-
trivial exponent $\epsilon_z\ne1$. In leading order we find that 
$\epsilon_z$ vanishes, and it next order it turns out to be negative, 
indicating  that the leading terms probably originate from 
analytic corrections and scale as 
$\langle S_z(\tau)S_z(0)\rangle \sim 1/\tau^2$. 
These results are, however, only 
approximate, and not very reliable, since the $\epsilon$-expansion formulas
are not applicable any more for $\epsilon_\perp=1$.

\subsubsection{Ising theory}

As we mentioned above, 
in  $\rm CeCu_{5-x} Au_x$ the magnetic degrees of freedom 
are more Ising-like. The corresponding quantum phase transition is therefore 
most probably be described by the Ising Bose-Fermi fixed point ($g_x = 
g_y=0$, $g_z\ne 0$, and $\lambda_z \ne \lambda_x=\lambda_y$)
within a dynamical mean field theory. This fixed point 
leads out of the region of applicability of our $\epsilon$ expansion. 
Nevertheless, from the Ward identity it still follows that 
\begin{eqnarray}
\epsilon_{z, \rm QCP}^{\rm UNI} \equiv 1\;,
\\
y_{h, z}^{\rm UNI} =1/2\;,
\end{eqnarray}
while the other exponents $\epsilon_\perp <1$ and $\eta_\perp>1$ are now 
determined by the specific form of the beta functions.

It is important to emphasize, that the prediction that for this locally critical 
dynamical mean field model local spin correlations in different directions decay 
with different powers and in one direction the decay is $\sim 1/\tau$ while in 
some other
Direction(s) it decays with a larger power, is rather robust, and it should be 
possible to test it by polarized neutron scattering.

\section{Basic definitions}

\subsection{Spectral representations}

All calculations presented here were carried out in Matsubara space
using the spectral representation of the various propagators.
Though it is not necessary for the calculations, we found it rather useful 
and intuitive to  construct  also  microscopic free Hamiltonians and fields
that  give rise to the local propagators in   Eqs.~(\ref{eq:<psipsi>}) 
and (\ref{eq:bose_korr}). 

We constructed the bosonic field $\phi$ in 
terms of one-dimensional bosons, $b_\alpha(k)$ as
\begin{equation}
\phi^\alpha \equiv \int_{-\Lambda}^\Lambda
dk\; |k|^{1-\epsilon\over 2} {1\over \sqrt{2}} 
( b_\alpha(k) +  b^\dagger_\alpha(-k))\;. 
\label{eq:phi_def}
\end{equation}
Here $b_\alpha(k)$ destroys a boson with polarization $\alpha$ and momentum 
$k$, and $\Lambda$ is the high-energy cut-off.
The $b_\alpha(k)$'s are normalized such that 
  $[b_\alpha(k),b^\dagger_\beta(k')]= 2\pi\; \delta_{\alpha\beta}\; \delta(k-k')$. 
Using a free bosonic Hamiltonian to generate their dynamics:
\begin{equation}
H^{(0)}_\phi \equiv \sum_{\alpha=1}^3 \int_{-\Lambda}^\Lambda
{dk\over 2\pi}\; |k|\; b^\dagger_\alpha(k) b_\alpha(k)\;,
\label{eq:H_0_bose}
\end{equation}
it can be readily verified that Eqs.~(\ref{eq:phi_def}) and (\ref{eq:H_0_bose})
lead to the propagator (\ref{eq:bose_korr}) with a constant 
$\Gamma(2-\epsilon)$. 
Note that $\phi$ is {\em not} the usual free bosonic field corresponding 
to Eq.~(\ref{eq:H_0_bose}) that would be defined through
\begin{equation}
\varphi^\alpha(x) \equiv \int_{-\Lambda}^\Lambda
dk\;  { e^{i k x} \over \sqrt{2 \;|k|}} \;
( b_\alpha(k)  +  b^\dagger_\alpha(-k))\;. 
\end{equation}

It is also a trivial matter to show that the  
bosonic Matsubara Green's functions can be represented as:
\begin{eqnarray}
{\cal D}^{(0)}(i\omega_n)& = & \int_{-\Lambda}^\Lambda
d\xi\; {\varrho(\xi) \over i\omega_n - \xi}\;, 
\label{eq:g_bose_free}
\\
\varrho(\xi) &=& {\rm sgn}(\xi) \; |\xi|^{1-\epsilon}\;,
\end{eqnarray} 
with $\varrho(\xi)$ the spectral function of the Bose field $\phi$.

Similarly, to represent the fermionic propagator we can use free 
one-dimensional chiral fermions, described by the Hamiltonian 
\begin{equation}
H^{(0)}_\psi  =  \sum_{\sigma=\pm} \int_{-\Lambda}^\Lambda
{dk\over 2\pi}\;  v_F k \; \psi^\dagger_{k,\sigma} \psi_{k,\sigma}\;,
\end{equation}
with $v_F \equiv  1$ the Fermi velocity. The fermion field $\psi^\dagger_{k,\sigma}$
creates a fermion with momentum $k$, spin $\sigma$, and energy 
$k$, and is normalized as $\{ \psi^\dagger_{k,\sigma}, 
\psi_{k',\sigma}\}  = 2\pi\;\delta(k-k')$. 
Then the correlation function of the field 
\begin{equation}
\psi_\sigma \equiv \int_{-\Lambda}^{\Lambda} {dk\over 2\pi} \; 
\psi_{k,\sigma}\;,
\end{equation}
behaves exactly as Eq.~(\ref{eq:<psipsi>}), and its Fourier transform 
can be represented as 
\begin{equation}
{G}^{(0)}(i\omega_n) =  \int_{-\Lambda}^\Lambda
{d\xi\over 2\pi} {1\; \over i\omega_n - \xi}\;.
\label{eq:g_fermi_free}
\end{equation}


\subsection{Pseudofermion technique}

In order to treat the spin dynamics in a field theoretical framework 
we used Abrikosov's pseudofermion technique.\cite{Abrikosov} 
In this approach one represents  the spin operator as 
$S^\alpha \to \sum_{m, m'}f^\dagger_m S^\alpha_{mm'} f_{m'}$, where $f_m$ 
is a pseudofermion annihilation operator corresponding to the $S^z = m$ 
spin component. In terms of these pseudofermions, the interaction terms 
can be treated using standard field theoretic methods.
However, in order to obtain a faithful representation of the spin 
one has to project out unphysical states with $\sum_m f^\dagger_m f_m > 1$.
This is achieved by adding a term
\begin{equation}
H_{\rm ps} = \mu_0 \sum_m f^\dagger_m f_m
\end{equation}
to the Hamiltonian, taking the chemical potential $\mu_0 \to \infty$, and 
keeping only the leading contributions  in the end of the calculation.

The propagation of free pseudofermions is described by the Matsubara 
Green's function
\begin{equation}
{\cal G}_{mm'}^{(0)}(i\omega_n) \equiv -\langle T  f_m  
f^\dagger_{m'}\rangle^{(0)}
_{i\omega_n} =  {\delta_{mm'} \; \over i\omega_n - \mu_0}\;.
\end{equation}

\section{Renormalization Group}

In order to analyze the behavior of the 
Bose-Fermi Kondo model at its non-trivial fixed points, we used field 
theoretical renormalization group methods, and applied expansion in the
small parameter $\epsilon$. 

\subsection{Spurious divergences and counterterms}

Simple power counting shows that
the pseudofermion self-energy contains 'spurious' 
contributions that diverge as $\sim \Lambda$.  These contributions 
only renormalize  the chemical potential, $\mu_0 \to \mu$, however, 
they lead to spurious divergences in higher order diagrams,
and must therefore be removed. 

To remove these divergences, we applied a {\em counterterm procedure}. 
In this approach we rewrite $H_{\rm ps}$ as
\begin{equation}
H_{\rm ps} = \mu \sum_m f^\dagger_m f_m + 
\delta\mu \sum_m f^\dagger_m f_m \;,
\end{equation}
and formulate the perturbation theory in terms of $\mu$ rather than 
$\mu_0$:
\begin{equation}
{\cal G}_{mm'}^{(0)}(i\omega_n)\to {\delta_{mm'} \; \over i\omega_n - \mu}\;.
\label{eq:g_ps_free}
\end{equation}
The  term $\sim \delta\mu \sum_m f^\dagger_m f_m $
must be treated as a perturbation and it 
 systematically cancels all spurious divergences in higher order 
diagrams. 
The coefficient $\delta\mu$ is  determined self-consistently
up to a given order of the perturbation theory
by the requirement:
$$
\Gamma^{(2)}\equiv {\cal G}^{-1}(i\omega_n=\mu) \equiv 0\;,
$$
and is given in leading order by
$$
\delta\mu = {S(S+1)\over 3} {\Lambda \over 1-\epsilon}
\sum_\alpha g_\alpha\;.
$$

\subsection{Vertex functions}

In order to compute the renormalization group equations, 
in addition to the inverse pseudofermion 
propagator $\Gamma^{(2)}$,  we have 
to compute the 'three-point' and 'four point' vertex 
functions, $\Gamma_\alpha^{(3)}$ and
$\Gamma_\alpha^{(4)}$  corresponding to the couplings $\gamma_\alpha$ and 
$\lambda_\alpha$ (see Fig.~\ref{fig:leading_corrections}). 
This is straightforward (though somewhat tedious)  within the pseudofermion technique. 

As we shall see later, the fixed point couplings at the non-trivial 
fixed points turn out to be small, $\gamma_\alpha \sim {\cal O} 
(\sqrt{\epsilon}) $ and  $\lambda_\alpha \sim {\cal O} (\epsilon)$.
This makes it possible to organize diagrams according to 
the powers of $\epsilon$ and perform a systematic expansion. 
The leading order diagrams are shown in Fig.~\ref{fig:leading_corrections}.

Having removed all 'spurious divergences' with the counterterm 
the diagrams contain only logarithmic infrared singularities 
$\sim {\rm ln}{\Lambda\over -\omega}$ in the $\epsilon\to 0 $ limit, 
and can be expanded as a power  of the logarithm. 
In leading order, the vertex functions are given by the 
following expressions:
\begin{eqnarray}
\Gamma^{(2)} &= &\omega \Bigl\{ 1 + {\rm ln}{\Lambda\over |\omega| }\; \;
\bigl[ {1\over 4}\sum_\alpha g_\alpha \bigr]\; 
+ \dots \Bigr\}
\label{eq:gamma2_leading}
 \\
\Gamma_x^{(3)} &= & \Lambda^{\epsilon/2} \gamma_x \Bigl\{
1 - {1\over4}(g_x-g_y-g_z) \nonumber
\\ &+& {\rm ln}{\Lambda\over |\omega |} \;\;\bigl[ {1\over4}(g_x-g_y-g_z) \bigr]\
+ \dots \Bigr\} 
\label{eq:gamma3_leading}
\\
\Gamma_x^{(4)} &= &  \lambda_x \bigl( 1 - {1\over4}(g_x-g_y-g_z)\bigr)
\nonumber \\
&+& {\rm ln}{\Lambda\over|\omega |}\;\;\Bigl[\lambda_y \lambda_z + 
{\lambda_x\over4} (g_x-g_y-g_z) \Bigr] + \dots 
\label{eq:gamma4_leading}
\end{eqnarray}
To perform the computation up to second order in $\epsilon$ 35 
diagrams must be taken into account. Their list and the corresponding 
rather lengthy results are given in the Appendix.

\subsection{The renormalization group transformation}

As mentioned above, 
once we removed all  spurious divergences, the perturbation 
series contains only {\em logarithmic divergences} in the $\epsilon\to 0 $
limit.  These logarithmic divergences can be systematically summed up 
and treated using renormalization group methods.\cite{Domb} 

The chemical potential $\mu$ has, of course, no physical meaning. 
Therefore, once the unphysical contributions corresponding to 
multiple occupancy of the $f$-states have been eliminated,\cite{footnote2}
we can transform it out by the transformation 
$i\omega_n \to \omega + \mu$.
After this transformation, all quantities depend exclusively on 
$\Lambda$, the dimensionless couplings, and the ratio $\omega/\Lambda$. 

To proceed further we define the pseudofermion $Z$-factor as
\begin{equation}
{\partial \Gamma^{(2)}\over \partial \omega}
\equiv Z^{-1}\bigl[\;{\omega/ \Lambda},\; \{q_i\}\;\bigr]\;,
\end{equation}
where $\{q_i\}$ ($i=1,..,6$) refers to the six dimensionless couplings
of the theory, and $\Gamma^{(2)}$ denotes the inverse pseudofermion
propagator.

\begin{figure}[t]
\centerline{\epsfysize 6.1cm 
{\epsffile{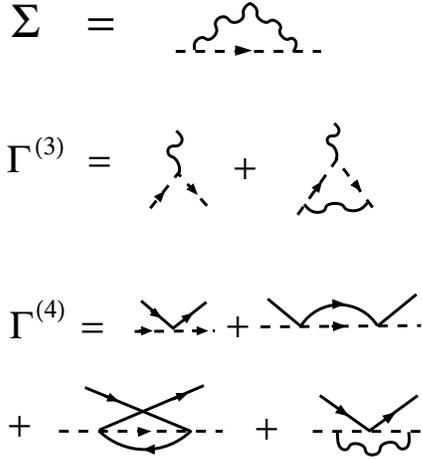}}}
\vspace{0.1cm}
\caption{
\label{fig:leading_corrections}
Leading order corrections for the pseudofermion self-energy and
vertex functions. Dashed  lines stand for free pseudofermion propagators
[Eq.~(\protect{\ref{eq:g_ps_free}})] 
wavy lines denote bosonic propagators [Eq.~(\protect{\ref{eq:g_bose_free}})] 
and continuous lines represent fermion  propagators
[Eq.~(\protect{\ref{eq:g_fermi_free}})].}
\end{figure}

Then the renormalizability of the theory implies that 
  there exists a transformation $\Lambda \to \Lambda'$, 
$q_i\to q_i'$ such that the 'renormalized' propagators and 
vertices remain invariant for all frequencies
up to ${\cal O}(\omega/ \Lambda)$ corrections:
\begin{equation}
\Gamma_{{\rm R} }^{(n)} \equiv
 Z \bigl(\;{\omega/ \Lambda}; q_i \bigr)\;
\Gamma^{(n)}\bigl(\;\omega, \Lambda; q_i \bigr) = \mbox{fixed} \;,
\label{eq:RGinv}
\end{equation}
where $\Gamma^{(2)}$ has been defined earlier, and  $n=3$ and $n=4$ 
refer to  the bosonic and fermionic 
vertex functions, respectively (see Fig.~\ref{fig:leading_corrections}).

This invariance is expressed through the
Callan-Symanzik equation:
\begin{equation}
{d\over dl} {\rm ln} \Gamma_{\rm R}^{(n)}
={\partial \over \partial l} {\rm ln} \Gamma_{\rm R}^{(n)}
+ \sum_i {d q_i\over dl}  {\partial \over \partial q_i} {\rm ln} \Gamma_{\rm 
R}^{(n)} = 0\;,
\label{eq:Callan}
\end{equation}
where we introduced the scaling parameter $l ={\rm ln}(\Lambda_0/\Lambda$), 
with $\Lambda_0$ the bare value of the cut-off.

Substituting Eq.~(\ref{eq:RGinv}) into Eq.~(\ref{eq:Callan}) we obtain
the following differential equations:
\begin{equation}
{\partial {\rm ln} \Gamma^{(n)}\over \partial l} +
{\partial {\rm ln} Z\over \partial l} 
+ \sum_\alpha {d q_\alpha\over dl}  
\left({\partial {\rm ln} \Gamma^{(n)}
\over \partial q_\alpha}  + {\partial {\rm ln}Z
\over \partial q_\alpha}\right) = 0\;.
\label{eq:gell_mann}
\end{equation}

To derive the scaling equations, we take the $\epsilon \to 0$ limit. 
Note that Eq.~(\ref{eq:gell_mann}) is satisfied for 
{\em all} frequencies and temperatures provided that 
they are much smaller than the cut-off $\Lambda$. 
 Therefore, expanding all vertex functions  in powers of 
${\rm ln}({\Lambda \over -\omega})$,  the Callan-Symanzik equations 
can be obtained by just picking the constant terms and 
the terms proportional to ${\rm ln}({\Lambda \over -\omega})$.  
[Of course, in a renormalizable theory 
 the same equations must be generated by comparing higher order terms in th expansion. Indeed,  as a check we verified in leading order in $\epsilon$ 
that the same scaling equations are obtained from comparing the 
$\sim{\rm ln}({\Lambda \over -\omega})$ and 
$\sim{\rm ln}^2({\Lambda \over -\omega})$ terms too.]  
Plugging in the expressions of  
Eqs.~(\ref{eq:gamma2_leading}-\ref{eq:gamma4_leading}) in the 
Eq.~(\ref{eq:gell_mann}) we obtain the scaling equations
Eq.~(\ref{eq:scaling_lambda_leading}) and 
Eq.~(\ref{eq:scaling_g_leading}). 
The second order ${\cal O}(\epsilon^2)$ scaling  
equations are given in the Appendix.

\subsection{Local magnetic field}

\subsubsection{Scaling equations}
To investigate the effect of local magnetic field we have to add the term 
Eq.~(\ref{eq:h_m}) to the Hamiltonian. As a consequence, 
${\cal G}^{(0)}$ is modified to:
\begin{equation}
 {\cal G}^{(0)} = {1\over i\omega - \mu + \sum_\alpha h_\alpha S^\alpha}\;,
\end{equation}
and $\Sigma$ also becomes a matrix in spin space. 
Restricting the calculation to the case where $h$ represents the smallest 
energy scale in the problem, $|h|< \omega, T$, we can expand
the renormalized inverse pseudofermion propagator as
$\Gamma^{(2)}_R = Z \; \Gamma^{(2)}$ as
\begin{equation}
\Gamma^{(2)}_R \equiv Z\;
[ \Gamma^{(2)}_0 + \sum_\alpha h_\alpha 
\Gamma^{(2)}_{1,\alpha} + {\cal O} (h^2) ]\;,
\end{equation}
where $\Gamma^{(2)}_0$ and Z denote the inverse propagator and $Z$-factor 
in the absence of  magnetic field, and $\Gamma^{(2)}_{1,\alpha}$ is also 
a matrix  in spin space:
\begin{equation} 
\Gamma^{(2)}_{1,\alpha} = - \Bigl.{\partial \Sigma \over \partial h_\alpha}
\Bigr|_{h=0}\;.
\end{equation}

From the invariance of $\Gamma^{(2)}_R$ it immediately 
follows that the magnetic field satisfies the following 
scaling equations:
\begin{equation}
{d\over dl}{\rm ln} \left[ Z h_\alpha \Gamma^{(2)}_{1,\alpha}\right] = 0\;.
\end{equation}
However, $\Sigma$ only depends on $h$ through the propagator
${\cal G}^{(0)}$. Therefore $\Gamma^{(2)}_{1,\alpha}$ is proportional
to the functional derivative ${\delta \Sigma/ \delta {\cal G}}$, 
and thus the bosonic vertex function by the Ward identity:
\begin{equation}
\Gamma^{(2)}_{1,\alpha} = - \Bigl.{\partial \Sigma \over \partial h_\alpha}
= \Lambda^{-\epsilon/2} {1\over \gamma_\alpha} \Gamma_\alpha^{(3)}\;.
\label{eq:ward}
\end{equation}

\begin{figure}[t]
\centerline{\epsfxsize 6.1cm 
{\epsffile{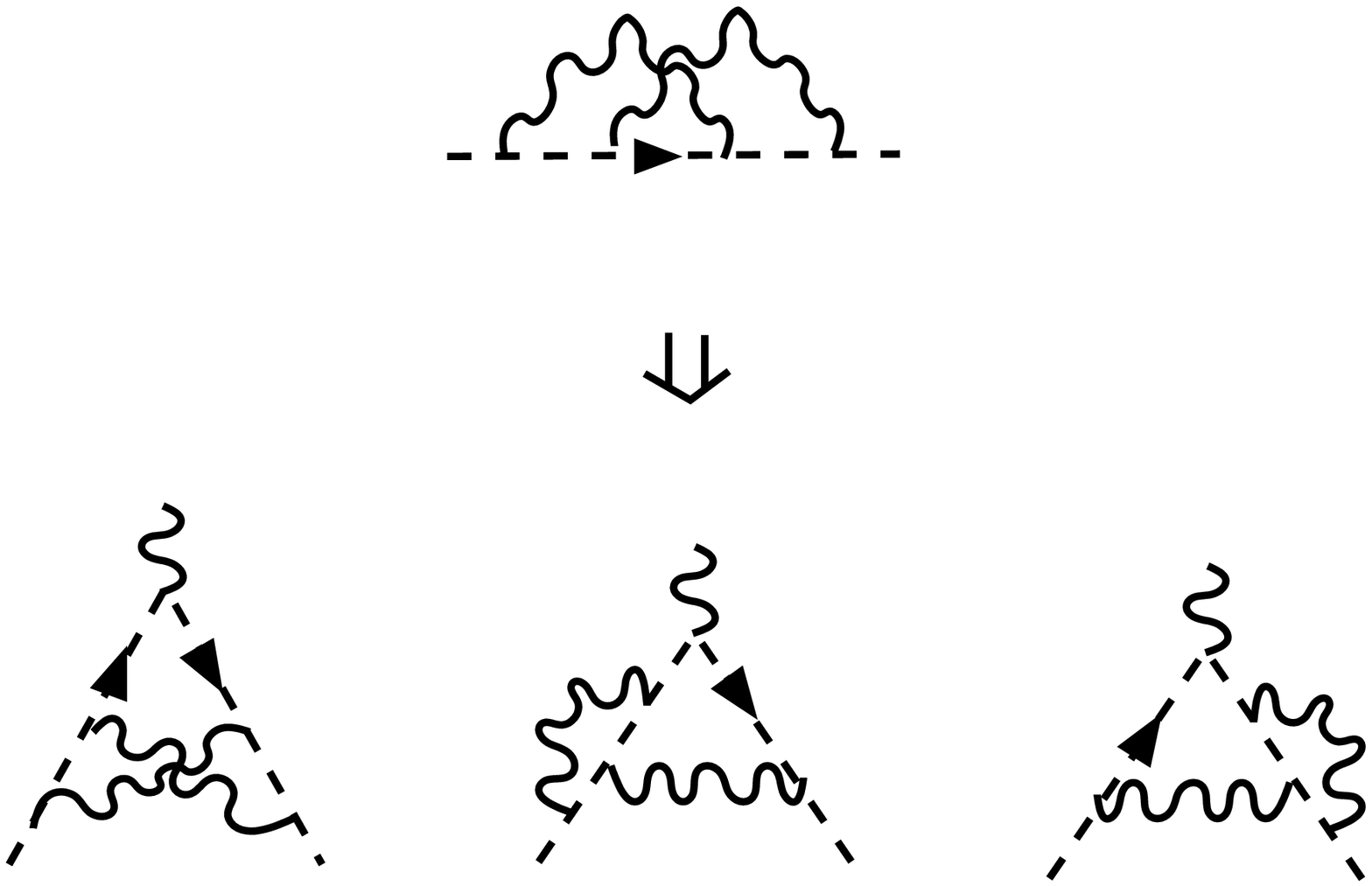}}}
\vspace{0.1cm}
\caption{
Diagrammatic representation of the Ward identity, 
Eq.~(\protect{\ref{eq:ward}}).
}
\label{fig:ward}
\end{figure}

Another way to derive this equation is to notice that ${\cal G}^{(0)}$ can 
be expanded as 
\begin{equation}
{\cal G}^{(0)} = {1\over i\omega - \mu} - \sum_\alpha 
h_\alpha {1\over i\omega - \mu} S^\alpha {1\over i\omega - \mu}
+\dots\;\,
\end{equation}
Differentiating  $\Sigma$ with respect to $h_\alpha$  at $h=0$ thus 
amounts in diagrams where one of the pseudofermion lines has been
replaced by an external vertex $\gamma_\alpha$ (see Fig.~\ref{fig:ward}).
It is easy to see that each vertex diagram is generated in this way exactly 
once and with the correct weight.
Combining the Ward identity Eq.~(\ref{eq:ward}) with the invariance of 
the renormalized vertex function, 
${d\over dl}{\rm ln}( Z \; \Gamma^{(3)})=0$, we immediately obtain 
that
\begin{equation}
{d\over dl}{\rm ln}\Bigl( {h_\alpha\over \Lambda^{\epsilon/2} \gamma_\alpha} 
  \Bigr) = 0\;,
\end{equation}
which, taking into account the definition of the $\beta$ function
leads to the scaling equation:
\begin{equation}
{1\over h_\alpha} {d h_\alpha\over dl} = 
-{\epsilon\over 2} + {1\over 2 \;g_\alpha} 
\beta^{(b)}_\alpha(\lambda_\alpha, g_\alpha)\;,
\end{equation}
or equivalently, Eq.~(\ref{eq:h_scaling}) for the dimensionless magnetic 
field, $\tilde h = h/\Lambda$.

\subsection{Susceptibility}

There are two 
possibilities at a non-trivial fixed point with couplings
 $g_\alpha = g_\alpha^*$ and $\lambda_\alpha = \lambda_\alpha^*$: 
\\
(a) If  $g_\alpha^*\ne 0$ then 
$\beta_\alpha^{(b)}(g_\alpha^*, \lambda_\alpha^*) = 0$ and
from Eq.~(\ref{eq:h_scaling})  it follows that 
\begin{equation}
\Bigl.{d {\tilde h}_\alpha \over dl}\Bigr|_{g^*, \lambda^*} = 
(1-{\epsilon\over2}){\tilde h}_\alpha\;.
\end{equation}
This immediately implies the exact result for the 
$\alpha$ component of the spin correlation function
at $T=0$:\cite{Cardy}
\begin{equation}
\langle T S^\alpha(\tau) S^\alpha(0)\rangle_{g^*, \lambda^*} = 
{1\over (T^*_\alpha \tau)^\epsilon}\;, \phantom{nnn}
\mbox{if $g_\alpha^*\ne0$},
\end{equation}
with $T^*_\alpha$ a dynamically generated energy scale
that is invariant under the scaling transformation,
similar to the Kondo scale.\cite{Hewson}

On the other hand, if $g^*_\alpha = 0$ then  
$\lim_{g^*_\alpha\to 0} \{
 \beta_\alpha^{(b)}(g_\alpha^*, \lambda_\alpha^*)/g^*_\alpha \} \ne 0$ 
takes some non-trivial value at the fixed point and no such universal 
relationship exists, and the spin-spin correlation function decays as
\begin{equation}
\langle T S^\alpha(\tau) S^\alpha(0)\rangle_{g^*, \lambda^*} = 
{1\over (T^*_\alpha \tau)^{\eta_\alpha}}\;, \phantom{nnn}
\mbox{if $g_\alpha^*=0$},
\end{equation}
with an exponent
\begin{equation}
\eta_\alpha = 
\epsilon - {\partial \beta_\alpha^{(b)}(g_\alpha^*, \lambda_\alpha^*) 
\over \partial g_\alpha^*}\;.
\end{equation}

The spin correlation functions decay as a power law only at $T=0$
temperature. It is, however, easy to extend these results to finite 
temperature by assuming that there exists a boundary conformal 
field theory\cite{Tsvelik} that corresponds to the critical dynamics of the 
fixed point. In this boundary conformal field theory the impurity
lives at the $x=0$ line of the complex plane, $z = \tau + i x$, 
and the theory is invariant under conformal mappings that 
 map this  boundary onto itself. 

The finite temperature correlation functions can be obtained by 
mapping the complex plane  on a strip of width $\beta=1/T$ using 
{\em e.g.} the function $w = {1\over \pi T} {\rm artg} \;z$. 
Assuming that $S_\alpha$ transforms as a primary field 
of dimension $\eta_\alpha/2$ 
\begin{eqnarray}
&&\langle S^\alpha(w_1) S^\alpha(w_2) \rangle_T =
\nonumber \\
&&\phantom{nnn}=
\Bigl({\partial z_1\over \partial w_1}\Bigr)^{\eta_\alpha/2}
\Bigl({\partial z_2\over \partial w_2}\Bigr)^{\eta_\alpha/2}
 \langle S^\alpha(z_1) S^\alpha(z_2) \rangle_{0}\;,
\end{eqnarray}
we obtain that at the critical point:
\begin{equation}
\chi_\alpha(\tau,T) = \Bigl({\pi T\over T_\alpha^*
{\rm sin}(\pi T \tau)}\Bigr)^{\eta_\alpha}\;.
\label{eq:chi_tau}
\end{equation}

Taking then the Fourier transform of Eq.~(\ref{eq:chi_tau})
and continuing it analytically to the real axis one obtains
an exact expression of the 
finite temperature  susceptibility:\cite{Tsvelik}
\begin{equation}
\chi_\alpha (\omega) = {T^\epsilon \over {T_\alpha^*}^\epsilon} 
{\pi^{\epsilon-1/2} \over T} 
{\Gamma({1\over 2} - {\eta_\alpha\over 2})
\Gamma({\eta_\alpha\over 2} - {i\omega \over 2\pi T}) \over 
 \Gamma({\eta_\alpha\over 2})
\Gamma({1\over 2} - {\eta_\alpha\over 2} - {i\omega \over 2\pi T})}\;.
\label{chi_full}
\end{equation} 
The scaling form Eq.~(\ref{eq:scaling_form}) readily follows 
from the asymptotic properties of the $\Gamma$ function.

\section{Conclusions}

In the present paper we analyzed the anisotropic Bose-Fermi Kondo problem 
using $\epsilon$-expansion methods. We found that in agreement with 
Ref.~\onlinecite{SenguptaBFKM}, spin anisotropy is {\em relevant} at 
the purely bosonic and Bose-Fermi  fixed points. 

We found two types of new bosonic fixed points:
One one of them has XY symmetry, and is unstable against breaking 
the XY symmetry. This fixed point may be relevant for 
non-magnetic impurities in antiferromagnets at their quantum critical point
 with easy plain anisotropy, and it may also describe the 
coupling of the spins to the Goldstone modes in an ordered 
phase of an antiferromagnet \cite{Antonio}.  The other fixed point has an 
Ising-like structure, and is stable. This fixed point could be relevant 
for non-magnetic impurities in an Ising antiferromagnet at the quantum 
critical point.

We identified furthermore two new Bose-Fermi quantum critical 
points in the impurity model 
with XY and Ising symmetries, respectively, which parallel 
the bosonic fixed points and separate the bosonic and fermionic 
quantum phases. We have shown that local spin correlations at these 
fixed points decay as $\langle T S_{x,y}(\tau) S_{x,y}(0)\rangle\sim 
1/\tau^\epsilon$ and $\langle T S_{z}(\tau) S_{z}(0)\rangle\sim 
1/\tau^\epsilon$ at the XY and Ising fixed points, respectively,
as guaranteed by a Ward identity. Correlations in the 
other directions decay faster with non-universal exponents that we 
computed to second order in $\epsilon$. These fixed points
may be relevant at the locally quantum critical points of
$\rm YbRh_2 Si_{2-x} Ge_x$ and $\rm CeCu_{5-x} Au_x$, 
respectively.\cite{Schroder,YbRh_2Si_2-xGe_x}

As discussed in detail in the Introduction, these results have 
very important implications for a possible dynamical mean field 
theory of the locally quantum critical behavior \cite{Si}: 
\\
(i) First of all, they 
imply that the real part of the local susceptibility diverges logarithmically
at the quantum critical point in the easy axis/easy plain directions, 
$\chi_{\rm easy, local}^{'}(\omega) 
\sim {\rm ln}(T^*/\omega)$, while the imaginary part 
is singular, $\chi_{\rm easy, local}^{''}(\omega) \sim {\rm sgn}(\omega)$.
In the other directions, however, the local susceptibility should show a 
power-law behavior,   $\chi_{\rm not-easy, local}^{'}(\omega)\sim 
\omega^\alpha$,  $\chi_{\rm not-easy, local}^{''}(\omega) \sim
\omega^\beta$,  and is likely to be dominated by analytic contributions
with $\alpha=0$ and $\beta=1$. These predictions should be directly observable 
by polarized neutron scattering and NMR relaxation measurements.
\\
{(ii)} A logarithmically divergent local susceptibility is  consistent 
with the  experimental scaling form, Eq.~(\ref{eq:c_exp}), only if the 
magnetic fluctuations are effectively {\em two-dimensional}. 
This quantum critical 
behavior can  therefore be only approximate, and if it applies to the 
systems above, then a cross-over to some other type of 
quantum critical behavior should occur at lower temperatures. 
\\
{(iii)} Finally, the above results imply an $h^2/T$ scaling for the local 
magnetic field. Whether  this is consistent with the experimentally observed
$h/T$ scaling with the global field, is an open question that needs to be 
answered. 

\begin{acknowledgements}
We are very grateful to P. Coleman, Kedar Damle, Gabi Kotliar, 
Subir Sachdev, and Q. Si\cite{Sidiscussions}  for stimulating discussions.
This research was supported by ITAMP, NSF Grants 
Nos.  DMR 99-81283 and DMR-0132874,
and Hungarian Grants No. OTKA F030041, 
and T038162. 
G.Z. is an E\"otv\"os  fellow.
\end{acknowledgements}

\vskip0.3cm
{\parindent=0pt {\em Note added:} After completing this work we learned that 
 a similar $\epsilon$ expansion has been carried out by L. Zhu and 
Q. Si with  very similar results.\cite{Siunpub}}

\section{Appendix}

In this appendix we give a few detailed expressions that 
may be important to reproduce or extend our results.

\begin{figure}[t]
\centerline{\epsfxsize 7.1cm 
{\epsffile{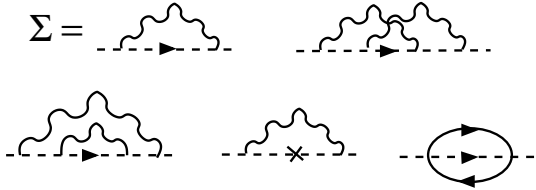}}}
\vspace{0.1cm}
\caption{
\label{fig:sigma_e2}
Second order corrections in $\epsilon$ for the pseudofermion self-energy.
The cross denotes the counterterm.}
\end{figure}

The easiest quantity to compute is the 
 pseudofermion self-energy, whose  diagrams are given 
in Fig~\ref{fig:sigma_e2} to ${\cal O}(\epsilon^2)$, 
and correspond to the  following expression:
\begin{equation}
\Gamma^{(2)} \approx \omega \Bigl\{
1 + 
\bigl ( 
{1\over4}\sum_\alpha ( g_\alpha + \lambda_\alpha^2)
-{1\over8}\sum_{\alpha\ne \beta} g_\alpha g_\beta + 
\bigr )
\;{\rm ln} {\Lambda\over -\omega} 
\Bigr\}
\;, 
\end{equation}
Note that the counterterm itself is also 
proportional to $\epsilon$ and is therefore important only 
in the $ {\cal O}(\epsilon^2)$ calculation.

\begin{figure}
\centerline{\epsfxsize 9.1cm 
{\epsffile{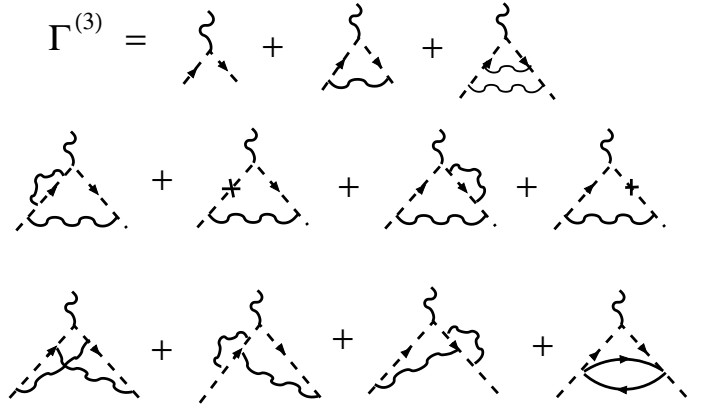}}}
\vspace{0.1cm}
\caption{
\label{fig:gamma3_e2}
Second order corrections in $\epsilon$ for the bosonic vertex function.}
\end{figure}

The expression of the bosonic vertex is somewhat lengthier:
\begin{equation}
\Gamma^{(3)}_\alpha  =  \Lambda^{\epsilon/2}
\gamma_\alpha \Bigl\{F_0^\alpha(g_\alpha,\lambda_\alpha)
+ {\rm ln} {\Lambda\over -\omega}F_1^\alpha(g_\alpha,\lambda_\alpha)
+ \dots \Bigr\}\;,
\end{equation}
where the coefficients $F_0$ and $F_1$ are given by
\begin{eqnarray}
F_0^x & = & 1 - {1\over 4} (g_x - g_y - g_z) + {\cal O}(\epsilon^2)\;, \\
F_1^x & = & {1-\epsilon\over 4}(g_x - g_y - g_z) 
- {1\over16}(g_x^2 + g_y^2 + g_z^2)
\nonumber \\
&+& {3\over 8} g_x (g_z + g_y) - {7\over 8 }g_y g_z
\nonumber \\
&+& {1\over 8 }(\lambda_x^2 - \lambda_y^2 -\lambda_z^2) + 
{\cal O}(\epsilon^2)\;.
\end{eqnarray}
The corresponding diagrams are depicted in Fig.~\ref{fig:gamma3_e2}.

\begin{figure}
\centerline{\epsfxsize 8.5cm 
{\epsffile{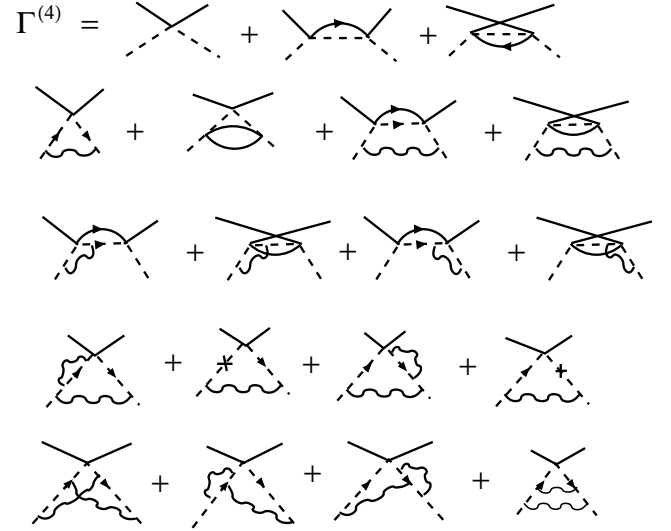}}}
\vspace{0.1cm}
\caption{
Second order corrections in $\epsilon$ for the fermionic vertex function.}
\label{fig:gamma4_e2}
\end{figure}

Finally, the  vertex corrections to the usual Kondo interaction 
to this order
are depicted in Fig.~\ref{fig:gamma4_e2}, and the corresponding expression 
reads:
\begin{equation}
\Gamma^{(4)}_\alpha = Q_0^\alpha(g_\alpha,\lambda_\alpha)
+ {\rm ln} {\Lambda\over -\omega}Q_1^\alpha(g_\alpha,\lambda_\alpha)
+ \dots \;,
\end{equation}
where $Q_0$ and $Q_1$ are given by
\begin{eqnarray}
Q_0^x & = & \lambda_x 
\bigl( 1 -  {1\over 4} (g_x -g_y - g_z) + {\cal O}(\epsilon^2) \bigr)\;, \\
Q_1^x & = & \lambda_y\lambda_z \bigl(  1 + {1\over 4} (g_y + g_z - g_x)\bigr)
+ \lambda_x \; F_1^x\;.
\end{eqnarray}
Here we only gave the $x$-component of the vertex functions, the others can be obtained by cyclic permutation.

The scaling equations can then be simply obtained by substituting 
these expressions into Eq.~(\ref{eq:gell_mann}) 
 and are given by
\begin{eqnarray}
{d\lambda_x \over dl} & = & \lambda_y\lambda_z 
- {1\over 2} \lambda_x(g_y+g_z) 
- {1\over 4}\lambda_x (\lambda_y^2 + \lambda_z^2)
\nonumber \\
& +& {1\over2}\lambda_x \bigl(g_x (g_y + g_z) - g_y g_z \bigr)
 + {\cal O}(\epsilon^4) \;,
\label{eq:beta^f_long}
\\
{dg_x \over dl}& = & g_x (\epsilon - g_z - g_y)
+ g_x^2 (g_y + g_z) 
\nonumber \\
&-& {1\over 2}g_x (\lambda_y^2 + \lambda_z^2)
 + {\cal O}(\epsilon^4) \;.
\label{eq:beta^b_long}
\end{eqnarray}

\draft\twocolumn[\hsize\textwidth\columnwidth\hsize\csname 
@twocolumnfalse\endcsname 
\begin{table}
\begin{tabular}{l|cc|ccc|c|c}
\baselineskip=2cm
 Fixed point   & $\{\lambda_\alpha\}$  &   $\{g_\alpha\}$  & 
$\Delta = -y_i$  & $y_r$ & $y_r^{\rm symm}$ 
& $R^{\rm imp}(T)$  
&  Stability \\  
\hline
Kondo $SU(2)$ & 
$\lambda_\alpha=\lambda \to \infty$ & $g_\alpha = 0$ & 
1  & none  & none &
$C_1 - C_2 T^2$ &
STABLE 
\\
$SU(2) Bose$ &
$\lambda_\alpha=\lambda\ne 0$ & $g_\alpha = 0$ & 
${\epsilon\over2} +{\epsilon^2\over8} $&
${\epsilon\over2} +{\epsilon^2\over2}$ &
stable &
$\sim T^{2\Delta}$ &
unstable 
\\
$SU(2)$ Bose-Fermi  &
$\lambda_\alpha=\lambda\ne 0 $ & $g_\alpha = g\ne 0$ & 
${\epsilon} - {\epsilon^2\over12} $ & 
${\epsilon\over2} +{7 \epsilon^2\over24}$ &
${\epsilon\over2} +{\epsilon^2\over 6} $ &
$C_1 + C_2 T^{\Delta}$ &
unstable 
\\
XY Bose &
$\lambda_\alpha = 0$ & $g_x = g_y\ne g_z=0$ & 
${\epsilon/ 2}$ &
${\epsilon} +{\epsilon^2} $ & 
stable &
$ T^{2\Delta}$ &
unstable 
\\
XY Bose-Fermi &
$\lambda_x = \lambda_y\ne\lambda_z $ & $g_x = g_y\ne g_z=0$ & 
${\epsilon - {9\epsilon^2\over 8}}$ &
${\epsilon} +{3\epsilon^2\over 4} $ & 
$ 0.618 \epsilon + 0.123 \epsilon^2 $ &
$C_1 + C_2  T^{\Delta}$ &
unstable 
\\
Uniax Bose-Fermi &
$ \lambda_x = \lambda_y\ne\lambda_z\sim 1 $ & 
$g_x = g_y = 0,\; g_z\sim1 $& 
? &
? & 
?  &
?  &
unstable
\\
Uniax Bose &
$\lambda_\alpha = 0$  & $g_x = g_y = 0,\; g_z\to\infty$ & 
? &
none & 
stable &
?  &
STABLE
\end{tabular}
\vskip0.2cm
\caption{
\label{fptable} 
Stability and structure of the various fixed points in the absence 
of magnetic field:  (a) $y_i$ is the dimension of the leading
irrelevant operator. This exponent governs thermodynamics and 
scattering rates at the fixed point. (b) $y_r$ is the dimension of 
the  leading relevant operator. This corresponds to breaking the 
symmetry of the fixed point. (c) $y_r^{\rm symm}$ is the scaling 
dimension  of the leading relevant operator within the subspace of 
given symmetry. This governs the quantum phase transition between 
the Bosonic and Fermionic Kondo fixed points.
}  
\end{table}
]
\narrowtext


\begin{references}
\bibitem{Jones}
I. Affleck, A.W.W.  Ludwig, and  B.A. Jones, 
Phys. Rev. B {\bf 52}, 9528 (1995). 
\bibitem{Sengupta2imp}
A. Georges and A.M. Sengupta, 
Phys. Rev. Lett. {\bf 74}, 2808 (1995).
\bibitem{Hofstetter}
W. Hofstetter and H.  Schoeller, 
Phys. Rev. Lett. {\bf 88}, 016803/1 (2002);
M. Vojta, R. Bulla, W. Hofstetter, Phys. Rev. B 65, 140405(R) (2002).
\bibitem{Avishai}
K. Kikoin and Y.  Avishai, Phys. Rev. Lett. {\bf 86}, 2090 (2001).
\bibitem{Schlottmann} P. Schlottmann, Phys. Rev. Lett. {\bf 84}, 1559 (2000).
\bibitem{Avi}
A. Schiller, F.B.  Anders, and D.L. Cox, 
Phys. Rev. Lett. {\bf 81}, 3235 (1998).
\bibitem{Koga}
M. Koga and D.L.  Cox, Phys. Rev. Lett. {\bf 82}, 2575 (1999).
\bibitem{Delft}
W. G. van der Wiel, S. De Franceschi, J. M. Elzerman, S. Tarucha, L. P. 
Kouwenhoven, J.   Motohisa, F. Nakajima, and T. Fukui, 
Phys. Rev. Lett. {\bf 88}, 126803 (2002).
\bibitem{Ye}
S. Sachdev and J. Ye, Phys. Rev. Lett. {\bf 70}, 3339 (1993).
\bibitem{Si96} Q. Si and J. L. Smith, Phys. Rev. Lett. {\bf 77},
3391 (1996).
\bibitem{Si99}
J. L. Smith and Q. Si, Europhys. Lett. {\bf 45}, 228 (1999)
\bibitem{SenguptaBFKM}
A.M. Sengupta, Phys. Rev. B {\bf 61}, 4041 (2000).
\bibitem{SubirBKM}
M. Vojta, C. Buragohain, S. Sachdev, Phys. Rev. B {\bf 61}, 15152 (2000);
S. Sachdev, M. Troter, and M. Vojta, Phys. Rev. Lett. {\bf 86}, 2617 (2001);
S. Sachdev, Physica C {\bf 357-360}, 78 (2001).
\bibitem{footnote} At a quantum critical point or in a d-wave 
superconductor the fermionic degrees of freedom may also have 
some anomalous dimension, but the discussion of this is out of the 
scope of the present paper.
\bibitem{Si}
Q. Si, S. Rabello, K. Ingersent, and L. Smith, 
Nature {\bf 413}, 804 (2001); for details see 
Q. Si, S. Rabello, K. Ingersent, and L. Smith, cond-mat/0202414.
\bibitem{Schroder} A. Schroder, G. Aeppli, R. Coldea, M.  Adams, 
O. Stockert, H.v.  Lohneysen, E. Bucher, R.  Ramazashvili, and P. Coleman,
Nature {\bf  407},   351 (2000).
\bibitem{Piers}
P. Coleman, C. Pepin, Q. Si, and R. Ramazashvili, 
Journal of Physics (Condensed Matter) {\bf 13}, 723 (2001).
\bibitem{SubirSG}
A. Georges, O. Parcollet, and S. Sachdev, Phys. 
Rev. Lett. {\bf 87}, 067202 (2001);
A. Georges, O. Parcollet, and S. Sachdev,
Physical Review B 63, 134406 (2001).
\bibitem{2CKM}
D.L. Cox and A. Zawadowski, Adv.  Phys. {\bf 47}, 599 (1998).
\bibitem{Hewson}
A.C. Hewson, {\em The Kondo problem to heavy fermions}
(Cambridge University Press, 1993).
\bibitem{Cardy}
John Cardy, {\em Scaling and renormalization in Statistical Physics} 
(Cambridge University Press, 1997).
\bibitem{Tsvelik}
M.C. Aronson, M.B. Maple, P. De Sa, A.M. Tsvelik, R. Osborn, 
Europhys. Lett. {\bf 40}, 245 (1997).
\bibitem{Kotliar}  
For a review on dynamical mean field theory see
A. Georges, G.  Kotliar, W. Krauth, and M.J.  Rozenberg, 
Rev. Mod. Phys. {\bf 68}, 13 (1996).
\bibitem{Kevin}
K. Ingersent and Q. Si, cond-mat/9810226.
\bibitem{Abrikosov}
A. A. Abrikosov, Physics {\bf 2}, 5 (1965).
 \bibitem{Domb} 
E. Brezin, J.C. Le Guillou, and J. Zinn-Justin, in {\em Phase 
transitions and critical phenomena-Vol. 6}, edited by 
C. Domb and M.S. Green (Page Bros. (Norwich) Ltd., 1996).
\bibitem{footnote2} 
The $\psi$ fermions' self energy in Fig.~\protect{\ref{fig:sigma_e2}}
is proportional to the thermal occupancy of the $f$-states in the 
$\mu\to \infty$ limit, $p=(2S+1) {\rm exp}\{-\mu/T\}$. To obtain the
{\em physical} self-energy one has to divide the one calculated with the 
pseudofermion technique by this quantity.
\bibitem{Antonio} A. Castro-Neto {\em et al.} (private communication).
In this work the purely bosonic and anisotropic problem 
is studied by a path integral approach yielding results very 
similar to ours. 
\bibitem{YbRh_2Si_2-xGe_x} 
O. Trovarelli, C. Geibel, S. Mederle, C. Langhammer, F.M. Grosche, 
P. Gegenwart, M. Lang, G. Sparn, and E. Steglich, Phys. Rev. Lett. 
{\bf 85}, 626 (2000).
\bibitem{footnote3} One can easily eliminate the $Z$-factor 
from Eq.~(\protect\ref{eq:gell_mann}) by subtracting the equation 
for $\Gamma^{(2)}$ from those of the vertex functions.
\bibitem{Siunpub} Q. Si (private communication).
\bibitem{Sidiscussions} 
We are especially grateful to Qimiao Si for discussions on the issue of 
$h/T$ scaling. 
\end{references}
\end{document}